\documentclass[aps,prl,reprint,superscriptaddress,longbibliography,showkeys,nofootinbib,floatfix]{revtex4-2}

\usepackage{graphicx}
\usepackage{amsmath}
\usepackage{amssymb}
\usepackage{physics}
\usepackage{xcolor}
\usepackage[T1]{fontenc}
\usepackage[hidelinks]{hyperref}
\begin{document}

\title{Neural Correlation Learning for Quantum-Enhanced Sensing with Time-Independently Driven Rydberg Atom Arrays}

\author{Tao Zhang}
\email{tao_zhang@mail.tsinghua.edu.cn}
\affiliation{Institute for Advanced Study, Tsinghua University, Beijing, 100084, China}
\affiliation{Intelligent Quantum Inception Co., Ltd., Haidian, Beijing, 100083, China}
\affiliation{iFLYTEK Research, Hefei, 230088, China}
\author{Xiaotian Nie}
\email{nxtnxt@mail.ustc.edu.cn}
\affiliation{Intelligent Quantum Inception Co., Ltd., Haidian, Beijing, 100083, China}
\affiliation{iFLYTEK Research, Hefei, 230088, China}
\author{Linghui Chen}
\email{lhchen@iflytek.com}
\affiliation{Intelligent Quantum Inception Co., Ltd., Haidian, Beijing, 100083, China}
\affiliation{iFLYTEK Research, Hefei, 230088, China}

\begin{abstract}
Quantum-enhanced sensing typically relies on preparing specific entangled states. However, engineering these states in scalable many-body systems remains challenging due to the difficulty of designing robust preparation protocols and accounting for realistic noise. We show that evolving a simple product state under native time-independent Hamiltonian dynamics in Rydberg atom arrays generates complex many-body correlations that can potentially enable quantum-enhanced parameter estimation reaching the Heisenberg limit (HL). The key challenge is then shifted to extracting the metrological information encoded in spatial correlations of measurement patterns. We introduce a neural correlation learning framework that combines Bayesian inference with neural networks trained on calibration data. Operating in a two-stage calibration-and-sensing protocol, we numerically demonstrate that this framework effectively extracts many-body correlations to saturate the Cram\'er--Rao bound dictated by the classical Fisher information. Consequently, it achieves quantum-enhanced sensitivity surpassing the standard quantum limit (SQL). Furthermore, the learned estimator is robust to realistic noise, and remains accurate in the absence of rare measurement patterns. Our results establish a hardware-efficient and scalable paradigm for quantum sensing in interacting many-body systems without requiring engineered entangled states.
\end{abstract}


\maketitle

\section{Introduction}

Quantum sensing exploits quantum correlations to surpass classical precision limits~\cite{giovannetti2011advances,degen2017quantum,pezze2018quantum}. A common paradigm is to prepare specific entangled states such as Greenberger--Horne--Zeilinger (GHZ) states~\cite{hyllus2012fisher, yin2025fast} or spin-squeezed states~\cite{kaubruegger2019variational, zhang2024entanglement, block2024scalable}, and then expose the prepared probe to the target field. Recent experiments have realized programmable metrological probes, interaction-based readout, scalable spin squeezing with Rydberg-mediated interactions, and GHZ-type clock states~\cite{marciniak2022optimal,colombo2022time,bornet2023scalable,eckner2023realizing,franke2023quantum,cao2024multi}. These capabilities rely on carefully designed entangling dynamics and control sequences. Maintaining the metrological gain in the presence of decoherence and control errors places stringent demands on protocol design and, in some settings, motivates error-correction constructions~\cite{kessler2014quantum,dur2014improved}. Meanwhile, this conventional ``prepare-then-sense'' sequence implicitly assumes that the external field to be measured can be sufficiently shielded during probe preparation. In practice, target fields can be global or difficult to switch off cleanly, so that the unknown parameter unavoidably participates in the state-preparation dynamics.

In order to bypass the above two requirements of complex quantum control and perfectly shielded environments, we utilize the natural Hamiltonian dynamics of interacting many-body systems. Instead of artificially separating state preparation and phase accumulation, the system starts from a trivial product state and evolves under a constant Hamiltonian where the target field is continuously present, merging state preparation and sensing into a unified process. Interacting many-body dynamics can generate long-range correlations and substantial metrological gain without first targeting a particular probe state~\cite{hauke2016measuring,dooley2021robust,puig2025manybody,sarkar2025exponential}. Related experiments of many-body signal amplification have recently been demonstrated in solid-state spin ensembles and cold Rydberg gases~\cite{gao2025signal,wang2026rydberg}. Rydberg atom arrays provide an ideal testbed for this paradigm~\cite{saffman2010quantum,browaeys2020many}. In such systems, strong van der Waals interactions and programmable geometries enable rich many-body physics~\cite{bernien2017probing,ebadi2021quantum,scholl2021quantum,chen2023continuous}, while high-fidelity site-resolved readout provides access to full spatial measurement outcomes~\cite{evered2023high}. However, the resulting entangled states after evolution are structurally complex. The conventional collective readout observables, such as the total excitation number $N_r = \sum_i n_i$ (coarse-grained collective readout) that are commonly used for states with simple collective-spin structure, fail in extracting these complex spatial correlations, leading to a substantial loss of the encoded metrological advantage. Then, the central challenge becomes: how to extract the quantum-enhanced sensitivity encoded in these complex many-body states?

To address this, we leverage neural networks to extract quantum sensitivity from these complex many-body states. Data-driven approaches have previously been used to learn quantum estimators from calibration data~\cite{hentschel2010machine,nolan2021machine}, construct adaptive Bayesian policies~\cite{fiderer2021neural}, and optimize sensing protocols in an end-to-end manner with parameterized variational quantum circuits and neural estimators~\cite{maclellan2024end}. While these approaches demonstrate the versatility of machine learning in quantum sensing, they typically require active quantum control to prepare tailored, readily readable probe states, or assume a decoupled ``prepare-then-sense'' sequence. Here, we embrace the complex states generated by the time-independent native dynamics, where the target parameter is continuously present. The classical neural network directly processes the spatial correlation in the measurement outcome and learns the mapping from spatial configurations to parameter posteriors. Our approach bypasses the need for an optimized state-preparation pulse sequence while exploiting spatial correlations that coarse-grained observables discard.

Crucially, to realize this in a hardware-efficient manner, our protocol operates in a highly practical two-stage framework. During the first \textit{calibration} stage, the neural network is designed to be trained directly on experimental snapshots collected under controlled parameter settings. Because this empirical training data inherently contains physical imperfections (e.g., spontaneous decay or miscalibration), the neural network automatically acts as a noise-aware decoder. In the subsequent \textit{sensing} stage, the neural network performs Bayesian inference to accumulate posteriors from measurement outcomes at the unknown parameter. We numerically demonstrate that this approach effectively extracts many-body correlations to saturate the Cram\'er--Rao bound dictated by the classical Fisher information, achieving quantum-enhanced sensitivity well beyond the SQL. The learned estimator is also remarkably robust to experimental noise such as decoherence. Furthermore, by learning correlations among different tweezer sites, the neural decoder can generalize to unseen patterns, and maintains near-optimal precision even when rare measurement patterns are omitted from the training data.

\section{Results}

\subsection{System and sensing framework}

\begin{figure}[tbp]
\centering
\includegraphics[width=\linewidth]{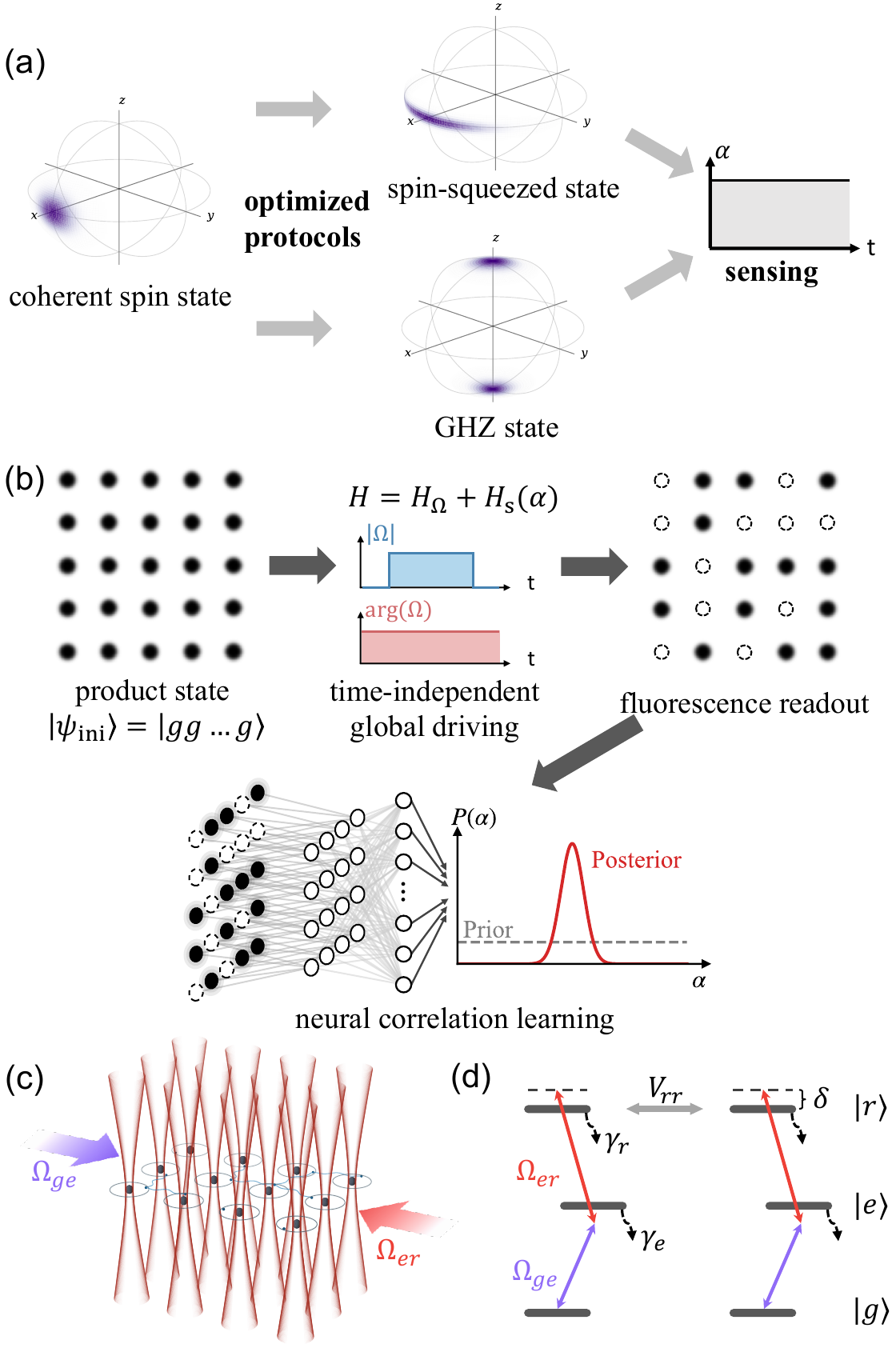}
\caption{\textbf{Neural correlation learning framework.} 
(a) Conventional quantum-enhanced sensing protocols prepare specific entangled states. The metrological gain of these states can be extracted with well-established protocols.
(b) Our protocol. A simple product state evolves under time-independent Hamiltonian dynamics where the target field is continuously present. The subsequent global fluorescence readout result is then sent into a neural decoder, which maps spatial correlations in the pattern to a posterior distribution over the unknown parameter $\alpha$.
(c) Schematic of neutral atoms trapped in a 2D tweezer array and driven by global Rydberg excitation lasers.
(d) Energy-level diagram. Two Rydberg beams coupling the ground state $|g\rangle$ to the Rydberg state $|r\rangle$ via an intermediate state $|e\rangle$ with Rabi frequencies $\Omega_{ge}, \Omega_{er}$, decay rates $\gamma_e$ and $\gamma_r$, two-photon detuning $\delta$, and Rydberg--Rydberg interaction $V_{rr}$.
}
\label{fig:illustration}
\end{figure}

\textit{Rydberg atom array.}---We consider $N$ atoms in a rectangular array with spacing $a$. The Rydberg excitation is implemented through the three-level ladder
$\ket{g}\leftrightarrow\ket{e}\leftrightarrow\ket{r}$ illustrated in Fig.~\ref{fig:illustration}(d). In the regime considered here, the intermediate state $\ket{e}$ is adiabatically eliminated, and the resulting dynamics within the $\{\ket{g},\ket{r}\}$ manifold is described by the effective two-level Hamiltonian
\begin{equation}\label{eq:hamiltonian}
H/\hbar = \sum_i \frac{\Omega_i}{2}\sigma_x^i - \sum_i \delta_i n_i + \sum_{i<j} V_{ij} n_i n_j,
\end{equation}
where $n_i=\ket{r}\bra{r}_i$ is the Rydberg-state number operator, $\sigma_x^i=\ket{g_i}\bra{r_i}+\ket{r_i}\bra{g_i}$, and $\Omega_i$ and $\delta_i=\delta_{\rm base, i}+\Delta$ are the effective two-photon Rabi frequency and detuning, respectively. The interaction strength is $V_{ij}=C_6/R_{ij}^6$, with $C_6$ the van der Waals coefficient. While our protocol can be directly generalized to inhomogeneous coupling cases, in the following we will focus on the homogeneous case ($\Omega_i=\Omega$, $\delta_i=\delta$) for simplicity. We take the amplitude of the external electric field $\boldsymbol{E}$ as the parameter of interest, which we aim to estimate through measurements. The presence of this external electric field induces a DC Stark shift on the highly polarizable Rydberg states. This shift leads to an additional, uniform energy offset to the Rydberg level, which directly maps to an increment $\Delta$ in the effective laser detuning. For the chosen Rydberg state, the relation between the detuning increment $\Delta$ and the electric-field amplitude $|\boldsymbol{E}|$ is determined by the Stark effect and can be calculated as $\Delta=\Delta_{\mathrm{Stark}}(|\boldsymbol{E}|)$~\cite{schlossberger2024rydberg,ocola2024control}. Thus, once this relation is specified, estimating $\Delta$ provides an estimate of $|\boldsymbol{E}|$.

Throughout the numerical calculations, we use $\Omega$ as the frequency unit and $R_0=(C_6/\Omega)^{1/6}$ as the length unit. Accordingly, detunings, interaction strengths, and decay rates are expressed in units of $\Omega$, time in units of $\Omega^{-1}$, and distances in units of $R_0$.

To perform the sensing task, the experimental protocol proceeds as shown in Fig.~\ref{fig:illustration}(b). In each experimental shot, the atom array is initialized in a trivial product state where all atoms are in the ground state, $\ket{\psi_{\mathrm{ini}}} = \ket{g, g, \dots, g}$. The system is then allowed to undergo coherent evolution under the native time-independent Hamiltonian in Eq.~(\ref{eq:hamiltonian}) for a fixed duration $t$, with the external field continuously present. Following the evolution, a projective measurement in the computational $Z$ basis is performed via site-resolved fluorescence imaging. Due to the anti-trapping nature of the optical tweezers for the Rydberg states, atoms in $\ket{r}$ are rapidly expelled from the traps and appear as dark sites, whereas atoms in $\ket{g}$ remain trapped and scatter fluorescence photons. This imaging process digitizes the many-body quantum state into a two-dimensional binary matrix corresponding to the array geometry, where we assign a bit value of $0$ to a fluorescent site (state $\ket{g}$) and $1$ to a dark site (state $\ket{r}$). Throughout this work, we refer to a specific binary snapshot obtained from a single experimental shot as a measurement \textit{pattern}. For a system of $N$ atoms, the total number of possible measurement patterns spans a space of $2^N$ patterns.

\textit{Fisher information framework.}---Measurements in the computational basis yield an occupation pattern $\mathbf{s} = (s_1,\ldots,s_N) \in \{0,1\}^N$ with probability $p(\mathbf{s}|\Delta)=|\langle\mathbf{s}\vert\psi(\Delta)\rangle|^2$. To evaluate the parameter sensitivity encoded in this probability distribution, we utilize the Fisher information framework. The classical Fisher information (CFI) quantifies the distinguishability of probability distributions of measurement outcomes under different parameter values. For the full pattern distribution, the CFI is
\begin{equation}
F_C^{\mathrm{pat}}=\sum_{\mathbf{s}}\frac{[\partial_\Delta p(\mathbf{s}|\Delta)]^2}{p(\mathbf{s}|\Delta)},
\end{equation}
while for the coarse-grained total Rydberg excitation number $N_r = \sum_i s_i$, it is $F_C^{N_r} = \sum_{m} (\partial_\Delta P_m)^2/P_m$, where $P_m = \sum_{\mathbf{s}:\,\sum s_i = m} p(\mathbf{s}|\Delta)$. For an unbiased estimator based on $M$ independent
measurements, the quantum Fisher information (QFI) sets the ultimate precision limit via the quantum Cram\'er--Rao bound~\cite{braunstein1994statistical,paris2009quantum}:
\begin{equation}
\mathrm{Var}(\hat\Delta) \geq \frac{1}{M F_Q},
\end{equation}
where $F_Q$ denotes the QFI associated with the estimation of $\Delta$. Here in our settings, at a fixed evolution time, the QFI scales with system size $N$ as at most $F_Q\propto N^2$, corresponding to the HL scaling. In the absence of interparticle correlations, the QFI is additive and scales at most linearly with $N$, corresponding to the SQL.

\textit{Bayesian parameter estimation and evaluation.}---To perform parameter estimation, we discretize the parameter range into $N_{\mathrm{bins}}$ grid points $\{\Delta_j\}$ uniformly distributed across the range of interest $\Delta \in [\Delta_{\min}, \Delta_{\max}]$. In the calibration stage, $M_{\mathrm{cal}}$ measurement shots at each $\Delta_j$ are sampled to train the neural-network estimator. During the sensing stage, the sensor is operated at an unknown detuning $\Delta_{\mathrm{true}}$. Given a sequence of $M$ measurement outcomes
$\mathbf{s}^{(M)}=\{\mathbf{s}_1,\ldots,\mathbf{s}_M\}$ that are
conditionally independent for a fixed $\Delta$, let
$P_k(\Delta_j)\equiv
P(\Delta_j|\mathbf{s}_1,\ldots,\mathbf{s}_k)$ denote the posterior
after the first $k$ measurements, with
$P_0(\Delta_j)=P(\Delta_j)$ being the initial prior. Bayes' rule updates
the posterior after each measurement according to
\[
P_k(\Delta_j)
=
\frac{
p(\mathbf{s}_k|\Delta_j)P_{k-1}(\Delta_j)
}{
\sum_{\ell}
p(\mathbf{s}_k|\Delta_\ell)P_{k-1}(\Delta_\ell)
}.
\]
Conditional independence implies that the joint likelihood factorizes as
$p(\mathbf{s}^{(M)}|\Delta_j)
=\prod_{k=1}^{M}p(\mathbf{s}_k|\Delta_j)$.
Iterating the single-shot update therefore gives
\begin{equation}
\begin{aligned}
P(\Delta_j|\mathbf{s}^{(M)})
&=
\frac{1}{\mathcal{Z}}\,
P(\Delta_j)
\prod_{k=1}^{M}p(\mathbf{s}_k|\Delta_j),\\
\mathcal{Z}
&=
\sum_{\ell}
P(\Delta_\ell)
\prod_{k=1}^{M}p(\mathbf{s}_k|\Delta_\ell),
\end{aligned}
\label{eq:bayes_update}
\end{equation}
where we take the initial prior to be uniform,
$P(\Delta_j)=1/N_{\mathrm{bins}}$. We use the posterior mean
\[
\hat{\Delta}(\mathbf{s}^{(M)})
=
\sum_j \Delta_j
P(\Delta_j|\mathbf{s}^{(M)})
\]
as the point estimator. The connection between
the single-shot likelihood in Eq.~(\ref{eq:bayes_update}) and the output
of the calibrated neural network will be established later.

To rigorously quantify the performance of the estimator, we use the mean squared error (MSE), defined as
\begin{equation}
\begin{aligned}
\mathrm{MSE}
&=
\mathbb{E}_{\mathbf{s}^{(M)}}\!\left[
\left(
\hat{\Delta}(\mathbf{s}^{(M)})-\Delta_{\mathrm{true}}
\right)^2
\right] \\
&=
\mathrm{Var}_{\mathbf{s}^{(M)}}\!\left[\hat{\Delta}\right]
+
\left(
\mathbb{E}_{\mathbf{s}^{(M)}}[\hat{\Delta}]
-\Delta_{\mathrm{true}}
\right)^2.
\end{aligned}
\label{eq:mse}
\end{equation}
Here, the expectation is taken over independent measurement sequences $\mathbf{s}^{(M)}$ generated at the true detuning $\Delta_{\mathrm{true}}$. The two terms in the second line correspond to the estimator variance and the squared bias, respectively. For an asymptotically unbiased and efficient estimator, the bias vanishes and the MSE approaches the classical Cram\'er--Rao bound, $\mathrm{MSE}\rightarrow 1/[M F_C(\Delta_{\mathrm{true}})]$.

\subsection{Information advantage of full measurement patterns}

In the conventional prepare-then-sense paradigm, a probe state is first prepared using a carefully designed protocol, and is then exposed to the unknown field. Such a separation is difficult to maintain for static or slowly varying fields that cannot be completely shielded during probe state preparation. Here in our protocol, the same Hamiltonian is used both for generating many-body correlations and encoding the detuning shift $\Delta$, thereby naturally avoiding this technical difficulty.

We first ask whether the time-independent evolution in Eq.~(\ref{eq:hamiltonian}) can itself generate a useful metrological resource. To this end, we use the QFI as the metric, which provides a measurement-independent characterization of the sensitivity of a quantum state to the parameter. It gives the maximum Fisher information achievable over all possible measurements. For the evolved pure state $\ket{\psi(\Delta,t)}$, the QFI can be calculated as~\cite{braunstein1994statistical,paris2009quantum}
\begin{equation}
F_Q(\Delta,t)=4\left[
\langle\partial_\Delta\psi\vert\partial_\Delta\psi\rangle
-\left|\langle\psi\vert\partial_\Delta\psi\rangle\right|^2
\right].
\label{eq:qfi_pure}
\end{equation}
Here, $\ket{\partial_\Delta\psi}\equiv\partial_\Delta\ket{\psi(\Delta,t)}$. Numerically, we evaluate this derivative using the central finite difference
$\ket{\partial_\Delta\psi}\simeq
[\ket{\psi(\Delta+\epsilon,t)}-\ket{\psi(\Delta-\epsilon,t)}]/(2\epsilon)$, with $\epsilon=10^{-4}$.

\begin{figure}[tbp]
\centering
\includegraphics[width=\linewidth]{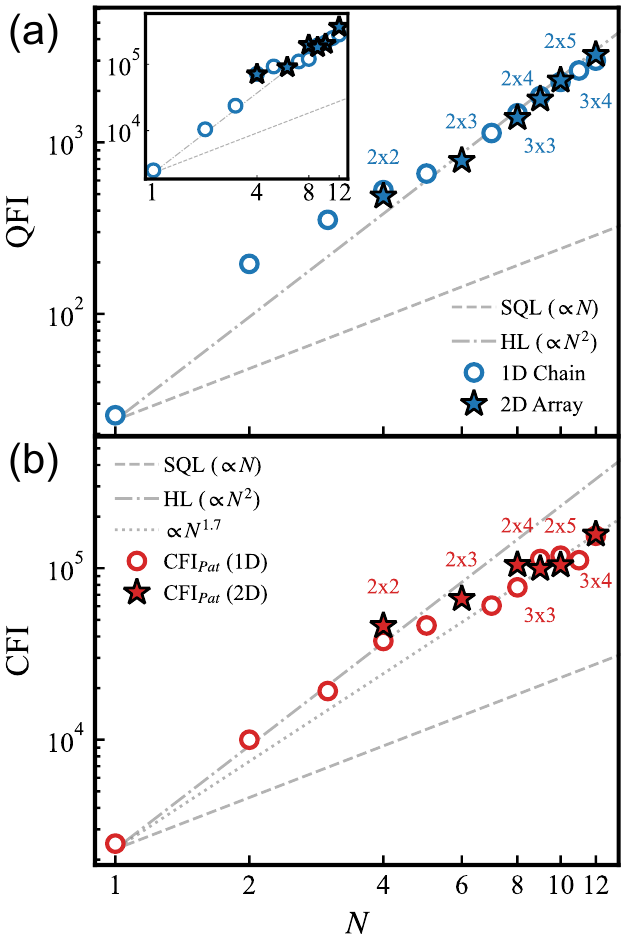}
\caption{\textbf{Scaling of metrological information.}
(a) QFI as a function of atom number $N$ for 1D chains and 2D arrays under time-independent Rydberg dynamics. Dashed and dash-dotted lines indicate the SQL scaling $\propto N$ and HL scaling $\propto N^2$, respectively. The main panel shows the QFI with fixed evolution time $t=10$, and the inset shows the QFI at $t=100$.
(b) CFI extracted from the full site-resolved measurement-pattern distribution at $t=100$. The dotted line shows the approximate scaling $F_C^{\mathrm{pat}}\propto N^{1.7}$. For each system size and geometry, the plotted value is optimized over spacing $a$ and base detuning $\delta_{\mathrm{base}}$, with fixed $\Omega=1$. 
}
\label{fig:qfi_cfi}
\end{figure}

Figure~\ref{fig:qfi_cfi}(a) shows the QFI under fixed $\Omega$ and evolution time, slightly optimized over the atomic spacing $a$ and base detuning $\delta_{\mathrm{base}}$ for each system size and geometry. For both one-dimensional chains and two-dimensional arrays, the QFI approximately follows the quadratic HL scaling over the system sizes considered. This behavior is observed at $t=10$ and persists at longer evolution time $t=100$ shown in the inset. These results demonstrate that time-independent interacting dynamics can indeed generate a metrologically useful many-body state without a separately engineered probe-preparation stage. The physical origin of the QFI enhancement can be understood from an equivalent generator representation. Defining $U_\Delta(t)=\exp[-iH(\Delta)t/\hbar]$, Eq.~(\ref{eq:qfi_pure}) can be written as
\begin{equation}
\begin{aligned}
F_Q(\Delta,t)
&=4\,\mathrm{Var}_{\psi_{\mathrm{ini}}}\!\left[G_\Delta(t)\right],\\
G_\Delta(t)
&=iU_\Delta^\dagger(t)\partial_\Delta U_\Delta(t)\\
&=\int_0^t d\tau\,
U_\Delta^\dagger(\tau)
\partial_\Delta\!\left(\frac{H}{\hbar}\right)
U_\Delta(\tau).
\end{aligned}
\label{eq:qfi_generator}
\end{equation}
For the Hamiltonian in Eq.~(\ref{eq:hamiltonian}) with the total detuning written as $\delta=\delta_{\mathrm{base}}+\Delta$, one has $\partial_\Delta(H/\hbar)=-\sum_i n_i$, such that $G_\Delta(t)$ is the time integral of the Heisenberg-evolved occupation operators. For independently evolving atoms initialized in the product state used here, these operators remain local and the variance of $G_\Delta(t)$ is additive, leading to the SQL scaling. In the presence of Rydberg interactions, their Heisenberg evolution involves support over multiple sites, producing intersite contributions to $\mathrm{Var}_{\psi_{\mathrm{ini}}}[G_\Delta(t)]$ and allowing the QFI to grow superlinearly with system size.

The QFI quantifies the maximum information available over all measurements. In a realistic implementation, one still has to access that information with a certain measurement strategy. We therefore evaluate the CFI of the complete site-resolved pattern distribution. This setting corresponds to a site-resolved fluorescence readout of all atoms, which is the most straightforward and native approach in neutral atom arrays. As shown in Fig.~\ref{fig:qfi_cfi}(b), the pattern CFI exhibits an overall superlinear increase after optimizing spacing $a$ and base detuning $\delta_{\rm{base}}$. Therefore, a substantial part of the quantum-enhanced sensitivity is maintained here by a direct computational-basis measurement. In the following discussion, we will take this specific measurement strategy as an example. However, we emphasize that further improvements in the CFI can be achieved by optimizing the measurement strategy, for example by applying local single-qubit rotation(s) before readout, and leave this as a future direction.

The above result reflects the spatially structured response generated by the interacting dynamics. Rydberg interactions imprint detuning-dependent correlations on the site-resolved measurement outcomes, such that the metrological information is distributed across the relative probabilities of many spatial configurations. Although the pattern CFI quantifies the information available in these outcomes, it does not by itself provide a practical estimator. Since the number of possible patterns grows exponentially with system size, explicitly reconstructing and evaluating the full likelihood becomes increasingly demanding. The remaining challenge is therefore to construct an efficient and experimentally feasible estimator for such a high-dimensional outcome distribution. This motivates the calibration-based neural-network estimator introduced below.

\subsection{Two-stage Bayesian sensing with a neural-network estimator}

\begin{figure*}[t]
\centering
\includegraphics[width=\textwidth]{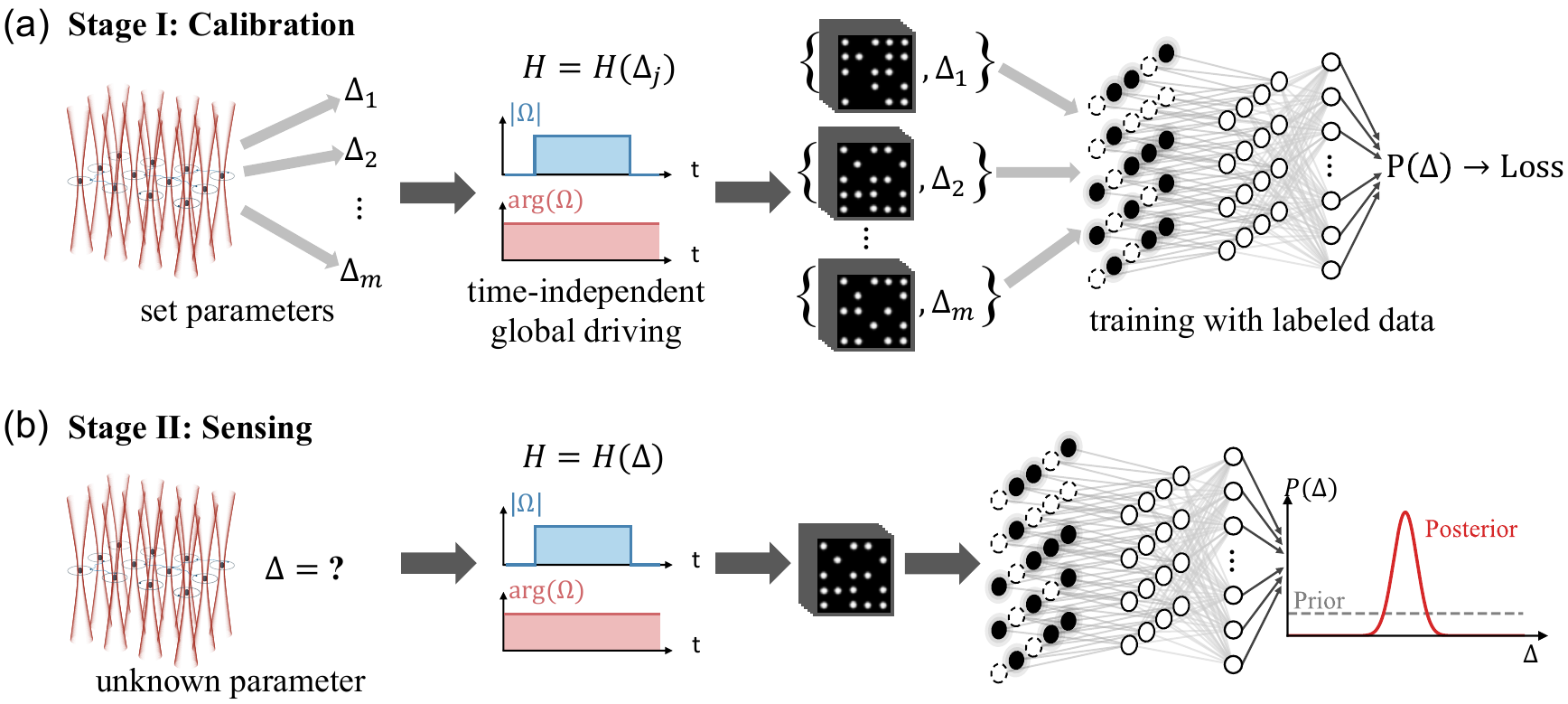}
\caption{\textbf{Two-stage calibration-and-sensing protocol.} 
(a) Calibration stage. The atom array is evolved under time-independent global driving at a series of known field values $\Delta_1, \Delta_2, \ldots, \Delta_m$. The resulting site-resolved binary measurement patterns are paired with their corresponding field labels to form a calibration dataset $\{ \text{pattern}, \Delta_j \}$. Training on these experimentally collected snapshots allows the neural network to learn the mapping from spatial correlations to parameter posteriors under the actual noise and hardware conditions.
(b) Sensing stage. The calibrated sensor is exposed to an unknown target field $\Delta=?$ and undergoes the same global driving dynamics. The single-shot measurement outcomes are fed into the trained neural network to obtain single-shot posteriors, which are recursively accumulated via Bayesian inference to produce a sharp posterior distribution $P(\Delta)$.
}
\label{fig:two_stage}
\end{figure*}

To this end, we introduce the two-stage calibration-and-sensing protocol illustrated in Fig.~\ref{fig:two_stage}. We infer the relevant pattern statistics directly from experimental calibration results and learn the spatial correlations with a neural network to perform pattern-to-posterior mapping. This setting naturally incorporates experimental imperfections that are present during both the calibration stage and sensing stage into the learned estimator, as will be discussed later.

During calibration, the sensor is operated at a set of known detuning shifts $\{\Delta_j\}$ spanning the sensing range of interest. At each calibration point, $M_{\mathrm{cal}}$ independent repetitions of the same time-independent evolution and site-resolved readout produce labeled snapshots $(\mathbf{s}_{k,j},\Delta_j)$. A neural network is then trained to associate each spatial pattern with a normalized probability distribution $\widehat P_\theta(\Delta_j|\mathbf{s})$ over the parameter values. The training is performed on single-shot measurements using the cross-entropy loss
\begin{equation}
\mathcal{L}_{\mathrm{CE}}
=-\frac{1}{N_{\mathrm{bins}}M_{\mathrm{cal}}}
\sum_{j=1}^{N_{\mathrm{bins}}}\sum_{k=1}^{M_{\mathrm{cal}}}
\log \widehat P_\theta(\Delta_j|\mathbf{s}_{k,j}).
\label{eq:ce_loss}
\end{equation}
Minimizing Eq.~(\ref{eq:ce_loss}) trains the network to approximate the single-shot posterior
$P_{\mathrm{cal}}(\Delta_j|\mathbf{s})$ associated with the calibration data. The likelihood factors required in Eq.~(\ref{eq:bayes_update}) can then be related to the network output via Bayes' rule:
\begin{equation}
\begin{aligned}
p(\mathbf{s}|\Delta_j)
&=
p_{\mathrm{cal}}(\mathbf{s})
\frac{
P_{\mathrm{cal}}(\Delta_j|\mathbf{s})
}{
P_{\mathrm{cal}}(\Delta_j)
}\\
&\simeq
p_{\mathrm{cal}}(\mathbf{s})
\frac{
\widehat P_\theta(\Delta_j|\mathbf{s})
}{
P_{\mathrm{cal}}(\Delta_j)
},
\end{aligned}
\label{eq:likelihood_from_network}
\end{equation}
where
$p_{\mathrm{cal}}(\mathbf{s})
=\sum_{\ell}p(\mathbf{s}|\Delta_\ell)
P_{\mathrm{cal}}(\Delta_\ell)$
is the marginal probability of observing $\mathbf{s}$ in the calibration data and is independent of the candidate detuning $\Delta_j$.

In the present calibration procedure, the same number of measurements $M_{\mathrm{cal}}$ is collected at each $\Delta_j$, giving the uniform calibration prior
$P_{\mathrm{cal}}(\Delta_j)=1/N_{\mathrm{bins}}$.
Equation~(\ref{eq:likelihood_from_network}) therefore reduces to
\[
p(\mathbf{s}|\Delta_j)
\simeq
N_{\mathrm{bins}}p_{\mathrm{cal}}(\mathbf{s})
\widehat P_\theta(\Delta_j|\mathbf{s}).
\]
The prefactor $N_{\mathrm{bins}}p_{\mathrm{cal}}(\mathbf{s})$ is independent of the candidate detuning $\Delta_j$ and therefore cancels upon normalization of the posterior. During sensing, the network parameters are fixed, and the same time-independent evolution and readout are repeated at the unknown target detuning $\Delta_{\mathrm{true}}$. Substituting the network
representation of the likelihood into the general Bayesian update in Eq.~(\ref{eq:bayes_update}) gives
\begin{equation}
P(\Delta_j|\mathbf{s}^{(M)})
\propto
P(\Delta_j)
\prod_{k=1}^{M}
\widehat P_\theta(\Delta_j|\mathbf{s}_k).
\label{eq:network_bayes_update}
\end{equation}
Thus, the information carried by successive independent measurement patterns is accumulated through their network-predicted single-shot posteriors. Note that here we have assumed the conditional pattern distribution $p(\mathbf{s}|\Delta_j)$ remains unchanged between calibration stage and sensing stage. This is the case for imperfections that are stable across these two stages, and will be discussed later. Details of the neural-network architecture and training procedure are provided in Methods.

\subsection{Sensing performance and precision benchmarking}

\begin{figure}[tbp]
\centering
\includegraphics[width=\linewidth]{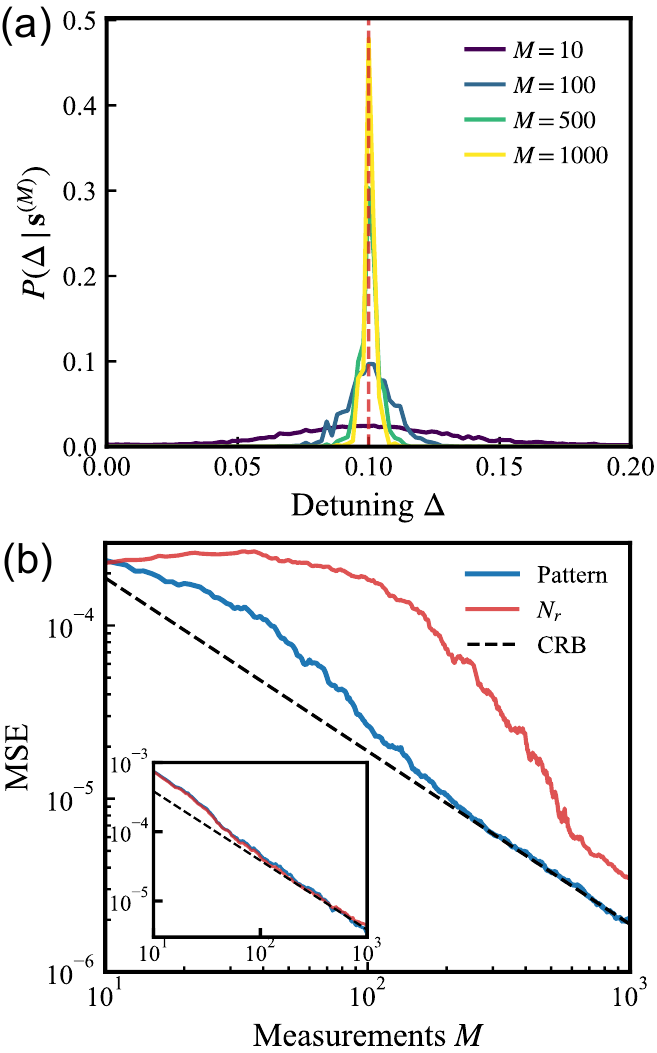}
\caption{\textbf{Sensing performance.} (a) Posterior distribution of the calibrated sensor tested at an unknown detuning, shown for increasing number of measurements $M=10$, 100, 500, and 1000 during the sensing stage. The posterior progressively sharpens around the ground truth (dashed line). Here, the system size is $2\times2$, and the number of calibration samples per bin is $10^3$. 
(b) MSE as a function of the number of measurements $M$ comparing pattern-based (blue) and $N_r$-based (red) Bayesian inference, with the pattern Cram\'er--Rao bound $1/[M F_C^{\mathrm{pat}}]$ (black dashed). Main panel: $3\times3$ array. Inset: $2\times2$ array. Results are averaged over 500 independent trials.}
\label{fig:performance}
\end{figure}

We next evaluate whether the learned neural-network estimator saturates the precision bound implied by the pattern CFI. For an ideal unbiased estimator, one would expect the posterior to be narrower while being centered on the true value as the number of experimental shots increases. Meanwhile, in order to build up connection with the Cram\'er--Rao bound, we adopt MSE defined in Eq.~(\ref{eq:mse}) as our evaluation metric, which provides a more stringent benchmark including both estimator variance and systematic bias. As an example, in Fig.~\ref{fig:performance}(a) we show the posterior distributions for a $2\times2$ array. For small $M$, the posterior remains broad because each measurement pattern carries only finite spatial correlation information about the unknown detuning. As $M$ increases, the posterior progressively narrows and becomes concentrated around the ground truth.
In Fig.~\ref{fig:performance}(b), we show that for both $3\times3$ and $2\times2$ (inset) arrays, pattern-based Bayesian inference converges toward the Cram\'er--Rao bound set by the pattern CFI and approaches the expected $1/M$ dependence. We also build a similar estimator based only on the coarse-grained $N_r$ observable. For a $2\times2$ array (inset), these two estimators give similar performance due to limited spatial correlation information at small system size. However, for a $3\times3$ array, the $N_r$-based estimator gives a substantially larger MSE. Both estimators are constructed from the same measurement snapshots and use the same Bayesian accumulation procedure. Their difference therefore directly reflects the information discarded when the site-resolved pattern is coarse-grained into the total excitation number $N_r$. Once patterns with different spatial structures are assigned the same value of $N_r$, the lost Fisher information cannot be recovered by subsequent classical processing.

Together, Figs.~\ref{fig:qfi_cfi} and~\ref{fig:performance} establish our whole sensing protocol: the time-independent native dynamics generates quantum states with a near-quadratically scaling QFI. Then, the native site-resolved measurement in the computational basis retains a superlinearly scaling CFI, and finally the calibrated neural-network estimator asymptotically extracts the metrological information. The quantum enhancement is therefore not confined to the QFI, but is converted into sensing precision beyond the SQL.

In the following two sections, we discuss some possible issues that may arise in realistic experimental implementations, and show that our protocol is robust to these imperfections.

\subsection{Robustness to experimental imperfections}

The precision benchmark above uses ideal theoretical models. In practice, experimental imperfections such as decoherence and miscalibration will modify the conditional pattern distribution $p(\mathbf{s}|\Delta)$. If these modifications are neglected during calibration, the resulting likelihood mismatch can introduce a systematic bias that does not vanish with increasing shot number $M$. A practical advantage of our calibration-based protocol is that imperfections which remain stable between calibration and sensing can be incorporated directly into the calibrated neural-network estimator, without requiring an accurate model of the nonideal dynamics.

As a representative example, we consider spontaneous decay from the Rydberg state $\ket{r_i}$ to the ground state $\ket{g_i}$. The open-system dynamics is described by the Lindblad master equation
\begin{equation}
\dot{\rho}
=
-\frac{i}{\hbar}[H,\rho]
+
\gamma\sum_i
\left(
L_i\rho L_i^\dagger
-\frac{1}{2}\{L_i^\dagger L_i,\rho\}
\right),
\label{eq:lindblad_decay}
\end{equation}
where $L_i=\ket{g_i}\bra{r_i}$ is the local jump operator, $\gamma$ is the spontaneous decay rate. We compare two calibration strategies under the same imperfect conditions. The ideal-model estimator is calibrated at $\gamma=0$ and subsequently applied to data generated at a nonzero decay rate. By contrast, the noise-aware estimator is calibrated using data generated under the same decay condition as the sensing process. This comparison quantifies the additional estimation error caused by likelihood mismatch, while the degradation that remains under matched calibration is consistent with a loss of information from the measurement patterns under open-system dynamics.

\begin{figure}[tbp]
\centering
\includegraphics[width=\linewidth]
{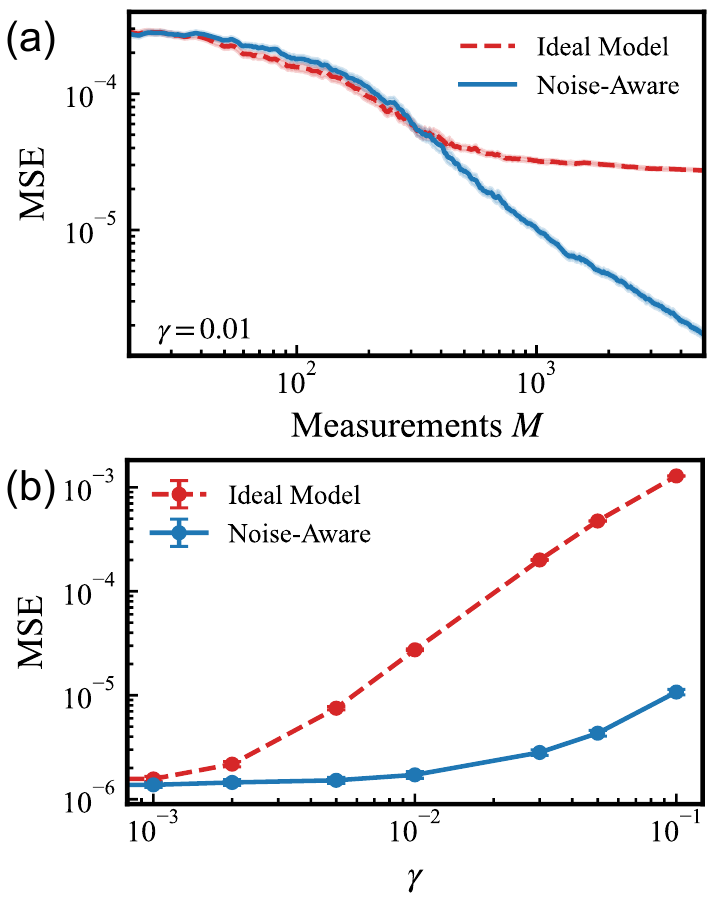}
\caption{\textbf{Noise-aware calibration under spontaneous decay.}
(a) MSE as a function of the number of measurements $M$ at $\gamma=0.01$. The ideal-model estimator uses a likelihood calibrated at $\gamma=0$, whereas the noise-aware estimator is calibrated at the same decay rate $\gamma=0.01$ as the sensing data. 
(b) MSE as a function of the spontaneous decay rate $\gamma$ for sensing sequences of length $M=5000$. All results are obtained for a $2\times2$ array and averaged over 500 independent sensing trials. For panel (b), the squared estimation error is first averaged within each trial over the final 100 Bayesian update steps, $M=4901,\ldots,5000$, before the ensemble mean and its standard error are evaluated across trials. Shaded regions and error bars denote the standard error of the mean.}
\label{fig:robustness_gamma}
\end{figure}

Figure~\ref{fig:robustness_gamma}(a) shows the MSE at the representative decay rate $\gamma=0.01$. At small $M$, both estimators are dominated by finite-shot fluctuations and exhibit comparable errors. As more measurement outcomes are accumulated, the noise-aware estimator continues to improve because the decay-modified pattern distribution is represented in its calibration data. At large $M$, likelihood mismatch produces a bias-dominated error floor for the ideal-model estimator, while the noise-aware estimator continues to improve as more sensing data is collected.
Figure~\ref{fig:robustness_gamma}(b) extends this comparison over the range of decay rates considered. The MSE of the ideal-model estimator increases rapidly with $\gamma$, indicating that even a moderate mismatch between the calibration and sensing distributions can dominate the estimation error at relatively large $M$. The noise-aware estimator remains substantially more accurate, although its MSE also increases at stronger decay. This remaining degradation is consistent with a physical loss of information from the measurement patterns caused by spontaneous decay. Although we take spontaneous decay as an example, the same principle can apply to many other experimental imperfections that are present during both the calibration stage and sensing stage such as miscalibration.

\subsection{Robustness to incomplete calibration patterns}

Another practical limitation arises from the finite size of the calibration dataset. For an array of $N$ atoms, the computational-basis measurement has $2^N$ possible outcomes. During the calibration stage, these outcomes are sampled according to the physical distribution $p(\mathbf{s}|\Delta)$. Consequently, rare patterns are usually undersampled, and some may be entirely absent from a finite calibration dataset. Whereas a pattern-by-pattern likelihood table provides no inference rule for such unseen outcomes, here we show that a neural-network estimator can generalize by exploiting spatial correlations shared among different patterns.

\begin{figure}[tbp]
\centering
\includegraphics[width=\linewidth]
{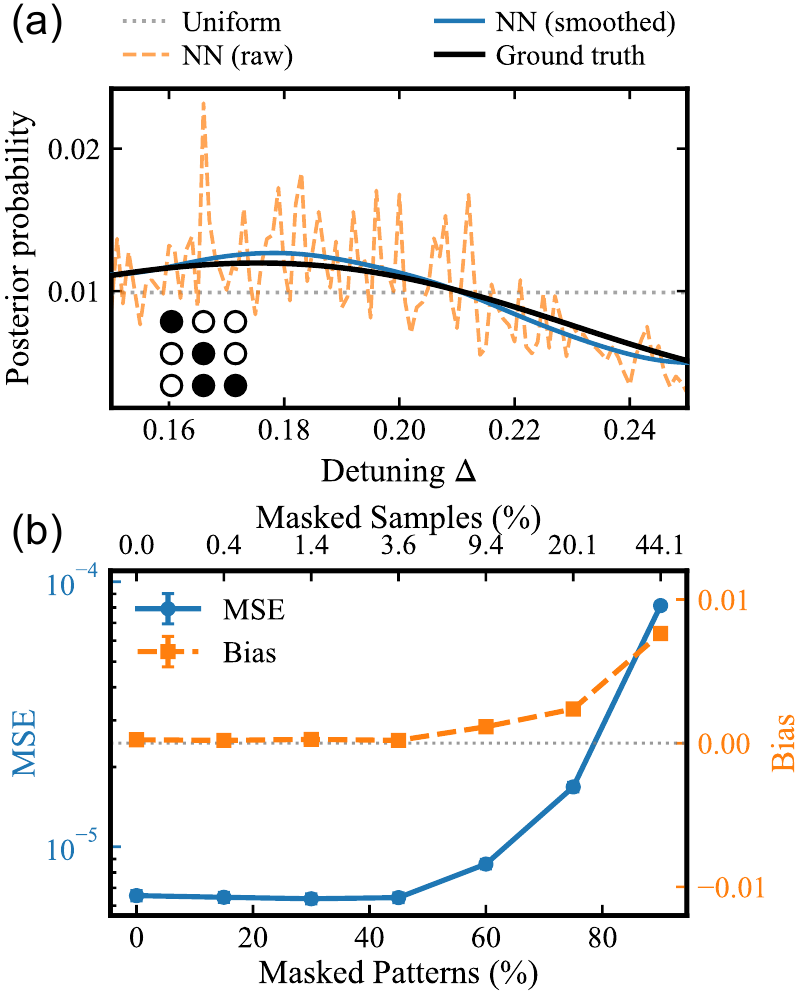}
\caption{\textbf{Robustness to incomplete calibration patterns.}
(a) Posterior distribution for an example pattern omitted from the training set. The gray dotted line denotes the uniform prior over the
discretized parameter grid. The exact posterior is compared with the raw neural-network prediction and its Gaussian-smoothed counterpart with a smoothing width of $\sigma=10$ grid points. Smoothing is used only to display the overall structure of this single-pattern prediction.
(b) Sensing performance as an increasing fraction of the least frequent patterns is removed from training. The bottom axis shows the fraction of masked pattern identities, while the top axis gives the fraction of calibration samples associated with those patterns. The MSE and estimator bias are evaluated using $M=1000$ sensing measurements and averaged over 1000 independent calibration-and-sensing realizations. All quantitative results in (b) use the raw, unsmoothed network outputs. Error bars denote the standard error of the mean across these realizations.}
\label{fig:robustness_mask}
\end{figure}

To test this capability, we consider a $3\times3$ array. As shown in Fig.~\ref{fig:robustness_mask}(a), we first exclude a single measurement pattern from the training set and compare the neural-network prediction for this unseen pattern with its exact posterior. Although the raw network output contains fluctuations, it captures the overall dependence of the posterior on $\Delta$. This agreement with the exact posterior can be more visible after a Gaussian smoothing, which is introduced only for this single-pattern illustration.

We next examine whether the estimator remains accurate when many patterns are absent from the training set. To construct this test, all patterns are ranked according to their occurrence frequency in the calibration data, and an increasing fraction of the least frequent patterns is removed before training. The sensing data are independently sampled from the complete physical distribution and may therefore contain these masked patterns. Figure~\ref{fig:robustness_mask}(b) reports the resulting MSE and estimator bias. The bottom axis denotes the fraction of masked pattern identities, whereas the top axis gives the fraction of calibration samples associated with those patterns. Here, masking the least frequent $45\%$ of all patterns removes only approximately $3.6\%$ of the calibration samples. Throughout this regime, the MSE remains nearly unchanged and the estimator bias remains close to zero. An obvious degradation appears only when the masked patterns account for an appreciable fraction of the sampled probability mass. Thus, we conclude that with such neural-network estimators, a large number of rare pattern identities can be absent from calibration without substantially affecting the final sensing precision. This stability under the removal of rare patterns suggests that the neural-network estimator exploits correlations shared across patterns, rather than operating solely as a pattern-to-posterior lookup table. We note that this result does not imply that the calibration cost is independent of system size, but shows that exhaustive enumeration of the $2^N$ outcome space is not necessary for the sensing task here.
\section{Discussion}

Our results show that, in an interacting Rydberg array, quantum-enhanced sensitivity can be achieved without preparing an entangled probe state with an optimized pulse sequence before the unknown parameter field is applied. Instead, the unknown target parameter participates in the same native time-independent evolution that generates useful many-body correlations, and a calibrated estimator recovers the information encoded in site-resolved measurement patterns. The metrological task is thereby shifted from constructing a specific entangled probe to inferring the parameter from the full distribution of experimentally accessible measurement patterns, which we address with a neural-network estimator. Our approach is particularly suitable for sensing static or slowly varying fields that cannot be completely screened during probe state preparation, and is robust to experimental imperfections that are present during both the calibration and sensing stage.

An immediate next step is to test this approach in arrays beyond the range of exact simulation. Since the estimator can be calibrated directly from device data, such an experiment does not necessarily rely on a classically calculated many-body likelihood. In this way, we can further determine how the accessible information and calibration cost scale with system size, and how frequently recalibration is required in the presence of experimental drift. Moreover, the Fisher information considered here characterizes only local sensitivity but not the full dynamic range. Adaptive adjustment of a global detuning offset or evolution time between experimental shots could broaden this range and track slowly varying imperfections, while preserving native time-independent evolution within each shot~\cite{hentschel2010machine,fiderer2021neural,belliardo2026multidimensional}. Simple local single-qubit rotations before readout may recover part of the remaining gap between the QFI and the pattern-based CFI~\cite{marciniak2022optimal}. Finally, site-resolved measurement patterns provide a natural starting point for multiparameter estimation of spatially varying fields~\cite{kaubruegger2023optimal}.

\section{Methods}\label{sec:methods}

\subsection{Numerical simulations}

The closed-system calculations were performed by exact diagonalization of the effective two-level Hamiltonian in Eq.~(\ref{eq:hamiltonian}) in the $2^N$-dimensional computational basis. Starting from the all-ground-state product state, we evaluated the evolved state through its spectral decomposition and obtained the pattern probabilities. We used the dimensionless units specified in the Results. The parameter derivatives in the calculation of QFI and CFI were evaluated by central finite differences with step $\epsilon=10^{-4}$; terms with pattern probability below $10^{-12}$ were omitted from the CFI division to avoid numerical singularities. For Fig.~\ref{fig:qfi_cfi}, the spacing and base detuning were scanned over 100 equally spaced values in $a\in[0.4,1.4]$ and 121 values in $\delta_{\mathrm{base}}\in[0,6]$, respectively, and the maximum Fisher information at the stated evolution time was retained for each size and geometry. During the optimization process, we find structured and regular parameter regions of enhanced Fisher information. For example, in the $N=2$ case, the optimal parameter region approximately follows the interaction-assisted two-photon resonance $2\delta_{\rm base}\simeq V(a)=C_6/a^6$
between $|gg\rangle$ and $|rr\rangle$. This resonance condition may therefore provide a physically motivated starting point for identifying optimal operating regimes.

For Figs.~\ref{fig:performance}--\ref{fig:robustness_mask}, we used $a=0.9$, $\Omega=1$, $\delta_{\mathrm{base}}=2.0$, and $t=10$. For Fig.~\ref{fig:robustness_gamma}, the open-system pattern probabilities were obtained by exponentiating the Liouvillian corresponding to Eq.~(\ref{eq:lindblad_decay}) and extracting the diagonal populations in the computational basis. Calibration and sensing samples were drawn independently from the resulting pattern distributions.

\subsection{Neural-network estimator}

The estimator was implemented as a fully connected feedforward network. The input of the network is the $N$-bit site-resolved pattern according to the system size, and the output is a normalized distribution over the 101 detuning bins using the softmax function. ReLU activations were used in the hidden layers. We used the Adam optimizer during training. The network was trained using the snapshot-level cross-entropy loss between the predicted detuning distribution and the labeled calibration data, as defined in the main text. For Fig.~\ref{fig:robustness_mask}, calibration snapshots sharing the same binary pattern were grouped into an empirical posterior over the detuning bins, and each retained pattern was treated as one training example. Neural network estimators were implemented and trained using the PyTorch framework. 

\section*{Data availability}
The numerical data supporting the results of this study are publicly available at \url{https://github.com/ZhangTao1999/Neural_Correlation_Sensing_Code_Data}.

\section*{Code availability}
The code used for the numerical simulations, neural-network training, and figure generation is publicly available at \url{https://github.com/ZhangTao1999/Neural_Correlation_Sensing_Code_Data}.

\section*{Acknowledgements}
The authors thank Intelligent Quantum Inception Co., Ltd. for providing the computational resources that supported the numerical simulations in this work.

\section*{Author contributions}
T.Z. conceived the idea, performed the numerical simulations, developed the software, analyzed the data, and prepared the figures. X.N. and L.C. provided supervision and guidance during the project. All authors contributed to discussions and the writing and editing of the manuscript.

\section*{Competing interests}
The authors declare no competing interests.

\bibliography{reference}

\end{document}